\documentclass[10pt,journal,compsoc]{IEEEtran}

\usepackage[T1]{fontenc}
\usepackage{courier}
\usepackage{upquote}
\usepackage{amsmath}
\usepackage{amssymb}
\usepackage{amsfonts}
\usepackage{graphicx}
\usepackage{booktabs}
\usepackage{hyperref}
\usepackage{cite}
\usepackage{tikz}
\usepackage{listings}
\usetikzlibrary{shapes.geometric, arrows, positioning}

\lstdefinelanguage{yaml}{
  keywords={true,false,null,y,n},
  keywordstyle=\color{blue},
  sensitive=false,
  comment=[l]{\#},
  morecomment=[s]{/*}{*/},
  commentstyle=\color{gray}\ttfamily,
  stringstyle=\color{red}\ttfamily,
  morestring=[b]',
  morestring=[b]"
}
\hypersetup{
    colorlinks=true,
    linkcolor=blue,
    filecolor=magenta,      
    urlcolor=cyan,
    pdftitle={Beyond Object Validation: Relational Conformance in Multi-Artifact Agent Releases},
    pdfpagemode=FullScreen,
}

\begin{document}

\title{Beyond Object Validation: Relational Conformance in Multi-Artifact Agent Releases}

\author{
    \IEEEauthorblockN{Tengjiao Liu} \\
    \IEEEauthorblockA{Founder \& Researcher, psi.run, \texttt{psi@psi.run}}
}

\markboth{Preprint, July 2026}%
{Liu: Beyond Object Validation}

\IEEEtitleabstractindextext{
\begin{abstract}
Agent systems validate inputs, tool calls, and generated objects. The final package often escapes the same scrutiny. In one DRSS release, the ledger supported 60 points and a failed certificate; the report announced a 100-point Gold Path. Every local gate was green. The package contradicted itself.

We study that failure alongside Schema Docs, where similar faults became product contracts, and Brand Shuttle GEO, where evidence is turned into repair work. The result is a candidate Schema-SIP Relational Conformance profile (SIP-RC). It models a release as a graph: claims point to evidence, decisions carry bounded authority, derived artifacts retain their execution conditions and lineage, and published bytes must match the package that was checked. Hard failures cannot be averaged away, and a validator recomputes critical decisions on a separate path. The paper establishes the failure class and shows that several mechanisms are practical. Whether the full profile performs better than existing checks remains an open experiment.
\end{abstract}

}

\maketitle
\IEEEdisplaynontitleabstractindextext
\IEEEpeerreviewmaketitle
\section{Introduction}

In July 2026, a DRSS development batch passed every automated, schema, and documentation gate. The release still contained a failed certificate under a Gold Path header, and a consumer report still carried manufacturing language. The consistency audit missed both. That batch did not test SIP-RC; it exposed the problem SIP-RC is meant to address.

The problem appears when an agent produces a package rather than one object. A research release joins evidence, findings, scores, certificates, and reports. A document exchange adds rendered views, assets, disclosure decisions, and hashes. A diagnostic product adds work orders and later review. Each part may satisfy its own schema while the package violates the relations that make it credible.

Schema Docs provides the working counterpoint. The local-first product handles Markdown, text, PDF, DOCX, PPTX, XLSX, and CSV. Real files exposed delayed spreadsheet headers, display-font semantics, Office formulas, unreachable assets, stale caches, mixed human/AI views, and unsafe outbound context. Those failures now live in executable contracts and regression tests.

Brand Shuttle GEO asks whether the same vocabulary survives outside research and document exchange. It maps public signals to a diagnosis, prioritized gaps, repair work, assets, and later review. We use it as an illustration, not as an independent validation.

Comparable boundary failures occur outside this research program. The PyTorch Foundation documented a nightly build that resolved a malicious \texttt{torchtriton} package from PyPI instead of the project index \cite{pytorch2022nightly}. EchoLeak showed how a crafted email could cross Microsoft 365 Copilot's trust boundaries and drive data exfiltration \cite{reddy2025echoleak}. Neither incident was analyzed with SIP-RC, but both confirm that package resolution and outbound disclosure can fail even when attention remains fixed on individual components.

The prior Schema Sandbox paper introduced a nine-layer constraint architecture, an L0–L3 classification, workspace partitions, and an early Schema Interoperability Protocol \cite{liu2026sandbox}. The present paper changes the unit of analysis from a sandbox or protocol object to the \textbf{released multi-artifact deliverable}.

\begin{figure}[htbp]
\centering
\begin{tikzpicture}[node distance=1.2cm, auto, >=latex', every node/.style={font=\sffamily\scriptsize}]
  \node [draw, rectangle, rounded corners, align=center, fill=gray!10] (E) {Evidence\\local check: PASS};
  \node [draw, rectangle, rounded corners, align=center, fill=gray!10, below=0.15cm of E] (C) {Certificate\\local check: PASS};
  \node [draw, rectangle, rounded corners, align=center, fill=gray!10, below=0.15cm of C] (P) {Report\\local check: PASS};
  \node [draw, rectangle, rounded corners, align=center, fill=gray!10, below=0.15cm of P] (A) {Assets\\local check: PASS};
  \node [draw, rectangle, rounded corners, align=center, fill=gray!10, below=0.15cm of A] (M) {Manifest\\local check: PASS};
  \node [draw, rectangle, align=center, right=1.0cm of P, fill=blue!5] (R) {Release\\graph};
  \node [draw, rectangle, align=center, right=1.0cm of R, fill=red!10] (X) {NON-CONFORMANT RELEASE\\score/status\\claim/evidence\\canonical/rendered\\asset/path\\approval/bytes};
  \draw[->, thick] (E) -- (R);
  \draw[->, thick] (C) -- (R);
  \draw[->, thick] (P) -- (R);
  \draw[->, thick] (A) -- (R);
  \draw[->, thick] (M) -- (R);
  \draw[->, thick] (R) -- (X);
\end{tikzpicture}
\caption{Local validity does not imply release conformance. The failure lies in relations among artifacts, not necessarily inside any one artifact.}
\label{fig:conformance}
\end{figure}
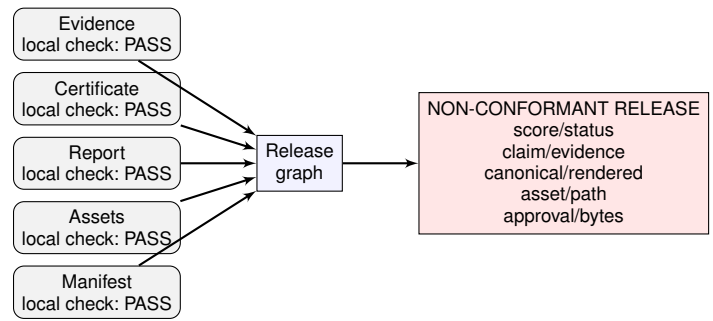

\subsection{From execution to release boundaries}

Earlier work in this research program treated persistence as a boundary problem, separated proposal from validation roles, and retained verified failures as constraints \cite{liu2026concretization, liu2026boundaries, liu2026unexpectedness, liu2026scars, liu2026cohorts, liu2026accommodation}. Schema Sandbox applied that line of thought to execution layers and workspace partitions \cite{liu2026sandbox}. DRSS exposed the missing boundary: a controlled run could still publish mutually inconsistent artifacts. Those prior studies explain the design lineage of SIP-RC; because they share an author and research program, they do not establish generality.

We ask three research questions:

- \textbf{RQ1:} Why did per-object validity and green local tests fail to guarantee deliverable conformance in the studied systems?
- \textbf{RQ2:} Which agent-specific relation and release obligations can be derived across the negative, successful, and transfer cases?
- \textbf{RQ3:} Relative to schema-only, shared-generator, and provenance-only baselines, what detection, false-block, authoring, and runtime costs does SIP-RC impose?

The case analysis answers RQ1 and proposes an answer to RQ2. RQ3 remains future work.

The profile covers claim support, authority, execution conditions, lineage, rendered artifacts, completeness, and publication—requirements derived from the incident record. Its guarantees stop at declared invariants. Section 7 describes the experiment still needed to test it.

Our working claim is:

> A multi-artifact deliverable should be accepted only after its release relations have been checked on a path separate from the generator's own acceptance logic.

\section{Related Work and Scope}

\subsection{Objects and graphs}

JSON Schema and constrained generation can keep an object well formed \cite{jsonschema}, \cite{geng2025jsonschemabench}, \cite{willard2023guided}. They cannot make its values correct \cite{singh2026structured}, much less reconcile it with another file. Data-quality systems already check cross-field consistency \cite{khare2019data}, and SHACL validates graph constraints \cite{shacl}, including planned and recursive forms \cite{figuera2021travshacl}, \cite{bogaerts2021recursive}. SIP-RC builds on that foundation with terms drawn from agent releases: evidence, decision authority, execution conditions, rendered views, hard gates, and publication state.

\subsection{Provenance, binding, and claims}

PROV-O records derivation and responsibility \cite{provo}. in-toto binds supply-chain steps to attestations \cite{intoto}, while C2PA binds manifests and assertions to assets \cite{c2pa}. SPDX and CycloneDX inventory software or system components \cite{spdx}, \cite{cyclonedx}. Pramāṇa comes closest to the present problem because it defines replayable, typed claim attestations \cite{kadaboina2026pramana}. All of these records can appear in a SIP-RC graph. The release decision still has to ask whether the evidence, verdicts, views, approvals, and final bytes agree.

\subsection{Runtime controls and release controls}

MemGPT organizes persistent context \cite{packer2023memgpt}. AgentSpec, SkillGuard, AC4A, ChainCaps, AIRGuard, and Agentproof govern runtime rules, permissions, composed authority, or workflow properties \cite{wang2026agentspec}, \cite{pan2026skillguard}, \cite{sharma2026ac4a, jiang2026chaincaps, qin2026airguard, xavier2026agentproof}. NeMo Guardrails and Guardrails AI check interaction stages or individual outputs \cite{nemo}, \cite{guardrailsai}. Those systems govern a run. SIP-RC checks the package left behind by the run.

\subsection{Interoperability boundary}

The Model Context Protocol connects hosts, clients, servers, tools, resources, prompts, and negotiated capabilities \cite{mcp}. A SIP-RC manifest can travel through MCP or another transport and can carry SHACL, provenance, or attestation records. The name needs one qualification: “SIP” already denotes the IETF Session Initiation Protocol \cite{rfc3261}. We therefore use \textbf{Schema-SIP} for the family and \textbf{SIP-RC} for this release profile.

\subsection{Comparative boundary}

\textbf{Table 1.} Where SIP-RC sits among adjacent mechanisms.

\begin{table*}[t]
\caption{Where SIP-RC sits among adjacent mechanisms.}
\label{tab:comparison}
\centering
\begin{tabular}{p{2.8cm}p{3.5cm}p{6.5cm}p{4.2cm}}
\toprule
\textbf{Approach} & \textbf{Primary strength} & \textbf{Gap motivating SIP-RC's additional layer} & \textbf{SIP-RC contribution} \\
\midrule
JSON Schema / structured output & Per-object structure and types & Cross-object agreement and rendered-package acceptance & Typed artifact relations and release predicates \\
SHACL / Trav-SHACL & General and efficiently planned graph constraints & Agent-specific decision authority, execution strength, and publication lifecycle & Agent release vocabulary and conformance profile \\
W3C PROV / in-toto & Provenance, derivation, supply-chain attestation & Business and epistemic release invariants & Release-gating semantics over provenance-bearing artifacts \\
C2PA & Claims, assertions, manifests, signatures, asset binding & Whether evidence and decision operators support an agent recommendation & Claim/operator/decision and completeness consistency \\
SPDX / CycloneDX & Component inventory and software/system supply-chain relations & Epistemic claims, decision authority, rendered views, and publication states & References BOMs while governing acceptance-relevant agent artifacts \\
Pramāṇa & Typed, replayable consequential-claim verification & Multi-file report/certificate/view/tier agreement & Package-level release conformance \\
NeMo Guardrails / Guardrails AI & Interaction-stage and individual-output validation & Frozen multi-artifact release agreement & Cross-output, cross-version release obligations \\
AgentSpec / Agentproof & Runtime rules and static/temporal workflow verification & Final multi-artifact product consistency & Artifact-centered release checks \\
SkillGuard / AC4A & Skill permissions and fine-grained resource access & Relation integrity of final conclusions and packages & Relational release profile \\
ChainCaps / AIRGuard & Composition-safe capability flow and action-time authority & Post-execution report/certificate/view/package agreement & Release-time authority and publication checks \\
MCP & Tool/resource connection and capability negotiation & Domain acceptance and release invariants & Complementary release obligations \\
\bottomrule
\end{tabular}
\end{table*}

The approaches reviewed here do not, by themselves, decide whether one agent release is fit to leave the system. SIP-RC groups release-specific rules—authority, hard gates, view agreement, typed approval, and publication state—on the same artifact graph.

\section{Relational Conformance Model}

The model says one thing precisely: a release is a graph, and validation looks for contradictions in that graph.

\subsection{Typed artifact graph}

A release contains files, views, approvals, and policies connected by relations such as “supports,” “renders,” and “authorized by.” We write that graph as:

\[
G = (V, E, \tau_V, \tau_E, \nu, A),
\]

where:

- \(V\) is a finite set of artifact nodes;
- \(E \subseteq V \times V\) represents directed semantic relations;
- \(\tau_V: V \to T_V\) maps nodes to an extensible type vocabulary that includes evidence, criterion, finding, recommendation, certificate, report, asset, rendered view, approval, policy, manifest, and package;
- \(\tau_E: E \to T_E\) maps edges to an extensible relation vocabulary that includes \texttt{supports}, \texttt{corroborates}, \texttt{evaluated\_by}, \texttt{summarizes}, \texttt{renders}, \texttt{references}, \texttt{derived\_from}, \texttt{authorized\_by}, \texttt{accepted\_by}, and \texttt{published\_as};
- \(\nu(v)\) identifies the immutable version and cryptographic hash of node \(v\);
- \(A \subseteq V\) is the authoritative node set from which release decisions must be recomputed.

Nodes may be mandatory, optional, unknown, blocked, or excluded from a partial release. A derived node may have multiple versions, but the release graph selects one immutable version for every included authoritative or rendered artifact.

\subsection{Object and release conformance}

Each node \(v_i\) has a local contract \(C_i\) validating object-level schema compliance. Each required relation or subgraph has a predicate \(R_j\) enforcing relational consistency. A predicate may span several edges—for example, recomputing one certificate from many findings—and one edge may participate in several predicates. The predicate count \(k\) therefore need not equal \(|E|\). Global release conformance is defined as:

\[
\begin{split}
\operatorname{ReleaseConform}(G) = {} & \left(\bigwedge_{i=1}^{|V|} C_i(v_i)\right) \land \left(\bigwedge_{j=1}^{k} R_j(G)\right) \\
& \land \operatorname{Publishable}(G),
\end{split}
\]

where \(k\) is the number of required relation or subgraph predicates declared by the active profile.

The DRSS counterexample establishes only the non-implication:

\[
\bigwedge_i C_i(v_i) = 1 \;\not\Rightarrow\; \operatorname{ReleaseConform}(G) = 1.
\]

\subsubsection{Minimal DRSS example}

The central D1 incident can be represented without the full case vocabulary:

\begin{scriptsize}
\begin{verbatim}
ledger(60) -eval(policy)-> dec(fail)
dec(fail)  -summarized---> cert(fail, 60)
cert       -rendered-----> report(Gold, 100)
\end{verbatim}
\end{scriptsize}

All four nodes pass their local schemas. Recomputing from \texttt{ledger + policy-v3} returns \texttt{fail, 60}, so the report relation emits \texttt{SIP\_RC\_ARTIFACT\_CONTRADICTION}. The validator blocks the report and its package. It may still release the ledger and failed certificate as a newly manifested partial graph if the audience is allowed to receive a failure record.

\subsection{Two-dimensional authority-scope orders}

One fixed chain cannot represent every domain's decision reach or publication extent. A profile therefore declares two finite partial orders:

\[
D=(S_D,\preceq_D),
\]

\[
B=(S_B,\preceq_B).
\]

For the cases in this paper, \(S_D\) contains observation, criterion, option, and project scopes; \(S_B\) contains field, object, artifact, bundle, release, and public-publication scopes. Their order is not assumed to be total: a profile declares which pairs are comparable and may add domain-specific scopes.

An authority scope is the product \(\sigma=(d,b) \in S_D \times S_B\), ordered component-wise:

\[
\sigma_1 \preceq \sigma_2 \iff d_1 \preceq_D d_2 \land b_1 \preceq_B b_2.
\]

A claim or approval \(c\) at scope \(\sigma_c\) can authorize an action \(a\) only when the action lies within the authority carried by that claim or approval:

\[
\operatorname{Authorized}(c,a) \iff \sigma_a \preceq \sigma_c.
\]

An aggregation or escalation operator may create a new authority token at a greater scope only when the profile declares the operator, its inputs, its approver role, and the hash-bound output. Without that new token, a criterion-level pass cannot authorize a project commitment, and approval of one artifact cannot authorize publication of an entire release. A violation is \textbf{reasoning- or release-scope escalation}, analogous to privilege escalation inside the deliverable graph. The model complements fine-grained access control \cite{sharma2026ac4a}, skill permissions \cite{pan2026skillguard}, and monotonic capability attenuation \cite{jiang2026chaincaps} without replacing their action-time enforcement.

\subsection{Execution risk and assurance}

Execution conditions mix two different questions. Risk records where and how the run acted; assurance records how well its result can be replayed:

\[
\begin{split}
\operatorname{risk}(m) & = (e,\mu), \\
\operatorname{assurance}(m) & = \rho,
\end{split}
\]

where \(e\) is externality (\texttt{local}, \texttt{pinned\_remote}, \texttt{open\_remote}), \(\mu\) is mutability (\texttt{read\_only}, \texttt{bounded\_write}, \texttt{arbitrary\_write}), and \(\rho\) is replayability (\texttt{none}, \texttt{trace\_only}, \texttt{deterministic\_replay}). Each profile declares the relevant comparisons. Ordinary derivation follows two rules:

\[
\begin{split}
\operatorname{risk}(m_{up}) & \preceq \operatorname{risk}(m_{down}), \\
\operatorname{assurance}(m_{down}) & \preceq \operatorname{assurance}(m_{up}).
\end{split}
\]

A downstream artifact must not under-report inherited risk or over-report assurance. Keeping the two orders separate avoids suggesting that greater replayability somehow cancels an external call or write.

An explicitly declared transform may strengthen replayability—for example, by adding a complete trace or deterministic fixture—but it cannot erase external calls or writes that already occurred. Its output must bind the transform version, inputs, retained trace, and validator result. Silent promotion, including relabeling fixture evidence as live or claiming deterministic replay without retained support, triggers \texttt{SIP\_RC\_MODE\_PROMOTION}.

\textit{Example.} Suppose a finding depends on \texttt{(pinned\_remote, read\_only, trace\_only)} evidence. A locally rendered report still inherits the \texttt{pinned\_remote} exposure in its release envelope; local rendering does not erase the remote observation. The report cannot claim \texttt{deterministic\_replay} unless an explicit transform retained sufficient fixtures and proved that stronger assurance. Conversely, a downstream step that performs an arbitrary write must declare that increased mutability even when all upstream artifacts were read-only.

\subsection{Non-compensatory gates}

A high score cannot hide a hard failure. If evidence is missing, a signature is invalid, or the package hash is wrong, the release fails even when every optional quality measure looks strong. Let \(g_k(G) \in \{0,1\}\) represent those hard gates, \(q(G)\) the quality score, and \(\theta\) its threshold:

\[
\operatorname{Accept}(G) =
\left(\bigwedge_k g_k(G)\right) \land (q(G) \ge \theta).
\]

The conjunction makes the rule explicit: quality matters only after every mandatory gate passes.

\subsection{Separate-path decision recomputation}

The generator does not grade its own work. It proposes \(\hat d\); a separately versioned validator reads the canonical ledger \(L \subseteq A\) and policy \(P\), computes \(d^*\), and compares the two:

\[
d^* = f_{validator}(L,P), \qquad \hat d = d^*.
\]

“Separate” means a distinct execution path, not automatically an independent implementation, author, or organization. The validator does not call the generator's acceptance-relevant helpers, decision templates, or scoring heuristics. Generic schema, parsing, or canonicalization libraries may be shared when they do not compute the decision and their versions remain explicit. A separately authored implementation, another language, or an external operator provides progressively stronger evidence and must be reported as such.

\subsection{Publication state machine}

The legal release states are:

\[
\begin{split}
\text{building} & \rightarrow \text{validated} \rightarrow \text{frozen} \\
& \rightarrow \text{verified} \rightarrow \text{published}.
\end{split}
\]

Only \texttt{published} is externally consumable. Mutation after \texttt{frozen} invalidates the candidate and returns it to \texttt{building}. A blocking violation found at or after \texttt{validated} does the same and invalidates endorsements, hash-mismatched approvals, and derived signatures. Publishing through a temporary directory followed by atomic rename prevents a receiver from observing a partial package.

\subsection{Evidence identity and corroboration}

Evidence deduplication and corroboration are different relations. Two observations with the same source identity and content hash resolve to one canonical snapshot; repeated ingestion does not create more support. Independently produced sources remain distinct evidence nodes linked by a corroboration relation. Collapsing both cases compromises source independence, whereas preserving every repeated snapshot inflates apparent support.

\subsection{Conditional conformance guarantee}

Let \(\mathcal{F}\) be the declared fault set encoded by mandatory SIP-RC predicates. Assume that (i) for every fault \(f \in \mathcal{F}\), the manifest declares every artifact and relation instance on which the predicate for \(f\) is defined—that is, the manifest vocabulary and instance graph subsume the faults the validator claims to detect; (ii) the validator correctly implements those predicates; (iii) signature and hash primitives behave as specified; and (iv) the published bytes equal the verified bytes:

> If \texttt{ReleaseConform(G) = true}, no correctly modeled and correctly implemented fault in \(\mathcal{F}\) is present in the published graph.

This target is conditional on the model and implementation. It offers no protection against undeclared faults, a bad domain method, or a bug shared by generator and validator.

\section{Study Method}

\subsection{Design and case roles}

We reconstructed three systems from repositories, artifacts, tests, and development records. Each system plays a different role:

The three systems share an author, a research program, and parts of their architecture. They are contrasting cases, not independent replications.

\subsection{Evidence and grading}

We examined repository code, manifests, schemas, generated artifacts, batch records, and development histories, and reran available local checks on 13 July 2026. Antigravity supplied a separately written account; we checked it against the files and use it only as internal triangulation.

- \textbf{E1:} fresh local re-execution or directly inspectable current artifact;
- \textbf{E2:} development log with observable failure and repair evidence;
- \textbf{E3:} current implementation or specification supporting a mechanism claim;
- \textbf{E4:} design intent or historical account not currently reproduced.

The Batch 16 DRSS document is a pending acceptance specification, not a completion report. It is used to identify observed defects and prescribed gates, never as proof that those gates passed.

The replication package containing all E1 execution logs, conformance vectors, and local schemas is included in the open-source repository.

\subsection{Evidence coverage}

Raw test totals appear once because the suites exercise different risks and are not comparable coverage scores.

\textbf{Table 2.} Evidence available in the three cases.

\begin{table*}[t]
\caption{Evidence available in the three cases.}
\label{tab:evidence}
\centering
\begin{tabular}{p{2.2cm}p{3.2cm}p{3.8cm}p{3.8cm}p{3.0cm}p{1.0cm}}
\toprule
\textbf{Case} & \textbf{Object checks} & \textbf{Cross-artifact checks} & \textbf{Render/package checks} & \textbf{Separate-path} & \textbf{External} \\
\midrule
DRSS & Historical Batch 13 reported 417/417 passing & Missing for the decisive score/status and audience contradictions & Consistency audit existed but missed both contradictions & No; generator path certified itself & No \\
Schema Docs & 267 passed, 0 failed, 1 skipped (268 total) & Implemented for canonical/view, hash, cache, disclosure, and receiver relations & Implemented across exchange packaging and a fourteen-step external-editing scenario & Implemented for selected checks; not a second product implementation & Receiver-side and desktop checks; no external replication \\
Brand Shuttle GEO & 116 passed, 0 failed, 2 skipped (118 total) & Implemented for selected evidence-mode and score-to-action contracts & Bilingual report and asset paths exercised; two live-network paths excluded & Focused mechanisms only & No \\
\bottomrule
\end{tabular}
\end{table*}

The last two columns matter more than the totals: none of the three systems has yet been replicated by an outside organization.

\subsection{Analysis}

For each included incident we reconstructed: input or precondition, expected relation, observed artifacts, local checks, relation failure, implementation status, candidate SIP-RC rule, and release action. Multiple assertions caused by one root failure were treated as one incident. Historical failure, injected fault, required repair, and completed repair were distinguished.

\section{Cross-Case Evidence}

\subsection{DRSS: central negative case}

DRSS evolved through repeated attempts to create an institutional research pipeline. The full batch chronology is retained as internal supplementary evidence; the main paper uses three standardized incidents.

\subsubsection{Incident D1 — Certificate/report contradiction}

\subsubsection{Incident D2 — Operator inversion}

\subsubsection{Incident D3 — Decision-scope escalation}

Two additional observations support, but do not complete, the protocol design. A package with zero completed dossiers and four gaps received a 100-point professional-report tier, motivating non-compensatory gates. A report changed after its manifest recorded bytes and hashes, motivating a freeze–verify–atomic-publish transaction. Both appear in a pending acceptance specification; their prescribed fixes are not counted as completed DRSS results.

\subsubsection{5.1.1 How the DRSS failures accumulated}

No single schema rule caused the DRSS release to fail. The damage accumulated in seven layers. Criteria, findings, metrics, and certificates were connected by array positions and substring matches instead of stable identifiers. Several components then summarized the same facts, leaving the package with more than one source of truth.

The next two layers concerned verification itself. Generator and validator shared helpers and domain assumptions, allowing the same error to appear on both sides of the check. Function-level tests then confirmed those shared paths without opening the final report and manifest as a single release. Passing tests were genuine, but their unit of observation was too small.

Three operational failures completed the pattern. Domain-specific compiler text, including battery units such as \texttt{GWh}, leaked into unrelated reports. Fixture, cached, and live evidence lacked explicit execution-mode labels. Specifications changed without version or hash bindings, leaving runtime rules and validators to diverge silently. Taken together, these seven layers explain why adding another object schema would not have repaired the release process; the checked object had to become the frozen artifact graph.

\subsection{Schema Docs: successful general-purpose interoperability}

Schema Docs implements a local-first workflow across Markdown, TXT, PDF, DOCX, PPTX, XLSX, and CSV:

\begin{scriptsize}
\begin{verbatim}
scoped ingest
  -> quality and fidelity diagnosis
  -> canonical representation
  -> human / AI / diagnostic views
  -> local selection and masking
  -> Send Gate
  -> exchange package and hashes
  -> receiver-side verification
  -> optional audited write-back
\end{verbatim}
\end{scriptsize}

Passing paths include conversion, package construction, content hashes, receiver/trust reports, timeline events, AI context, and a fourteen-step external editing scenario. Historical incidents are used here to explain how the successful contracts were discovered.

\subsubsection{Incident S1 — Stored value versus displayed meaning}

\subsubsection{Incident S2 — Asset existence versus receiver reachability}

\subsubsection{Incident S3 — Input identity versus derivation freshness}

Schema Docs also makes disclosure a product state. Local masking replaces detected secrets, email addresses, IP addresses, phone numbers, and configured PII classes with typed placeholders. A Send Gate returns \texttt{allow}, \texttt{review\_required}, or \texttt{block}, with reasons and next actions. Local SQL filtering can reduce CSV/XLSX context before Markdown handoff. Credential-like content stops at \texttt{review\_required}; the product will not send it from that state. The mechanism is limited, but it is visible and testable.

\subsubsection{5.2.1 What the product case establishes}

The product now enforces four rules that began as unrelated bugs: stored value must agree with displayed meaning; assets must resolve for the receiver; cached output must name the extractor that produced it; and send approval must bind the reviewed bytes. The rules still live in different parts of the codebase. SIP-RC proposes a common declaration for them.

\subsection{Brand Shuttle GEO: illustrative transfer case}

GEO maps a bounded set of public evidence to a diagnosis, prioritized gaps, repair work, assets, and recurring review. It lets us test the vocabulary on an evidence-to-action product. It cannot establish generality.

\subsubsection{Incident G1 — Degraded execution mode}

\subsubsection{Mechanism G2 — Evidence-to-action closure}

\subsubsection{5.3.1 From diagnosis to bounded action}

GEO ends with named deliverables rather than open-ended advice. Diagnostic scores produce repair work, update intervals, and localized assets. The input and output contracts can be public even when prompts, crawlers, and scoring heuristics remain private. That separation is useful for review, but GEO is still an internal case.

\subsection{Cross-case synthesis}

Across the cases, the same seven relations recur. Schema Docs implements several of them in a working product; GEO exercises a smaller subset.

\section{Candidate SIP-RC Profile}

\subsection{Scope and normative language}

SIP-RC is a candidate profile, not a standard. Its conformance requirements have not yet passed comparative evaluation. The profile applies to releases with at least two artifacts or views that affect acceptance. MUST, MUST NOT, SHOULD, SHOULD NOT, and MAY follow BCP 14.

SIP-RC extends rather than replaces SIP-Core and SIP-Secure:

\begin{scriptsize}
\begin{verbatim}
SIP-Core: object contracts + structured rejections
   ↓
SIP-Secure: signatures + capabilities + evidence
   ↓
SIP-RC: cross-artifact relations + release gates
\end{verbatim}
\end{scriptsize}

A SIP-RC implementation inherits the mandatory requirements of the lower profiles. Deployment and marketplace metadata may be defined by separate profiles, but they are outside this paper's release-conformance contribution.

SIP-RC does not require RDF. A conforming implementation MUST provide:

1. a typed artifact graph or equivalent relation model;
2. per-object schemas;
3. required relation predicates;
4. an immutable release manifest;
5. a validator report with stable violation codes;
6. a publication state and release verdict.

JSON Schema MAY validate objects. SHACL MAY validate an RDF/JSON-LD graph. A restricted Datalog or SIP-RC invariant DSL MAY express domain rules. Interoperability depends on shared semantics and test vectors, not on one mandatory rule engine.

\subsection{Required modules}

\subsection{Version and extension envelope}

\begin{lstlisting}[language=yaml]
protocol: "schema-sip"
version: "1.2-draft"
profile: "sip-rc"
status: "candidate"
namespace: "https://schema-sip.example/ns/rc/1.2"
extensions:
  - id: "org.example.domain-claims"
    version: "2.0.0"
\end{lstlisting}

Unknown mandatory extensions MUST cause rejection. Unknown optional extensions MAY be preserved without enforcement if the release manifest declares that they do not affect authoritative decisions.

\subsection{Minimal relation declarations}

\begin{lstlisting}[language=yaml]
claim_contract:
  semantic_scope: [observation, criterion, option, project]
  publication_scope: [field, object, artifact, bundle, release, public_publication]
  scope_order: "profile-declared partial orders"
  deny_implicit_scope_escalation: true
  require: [evidence, operator, execution_mode]

execution_mode:
  risk:
    externality: [local, pinned_remote, open_remote]
    mutability: [read_only, bounded_write, arbitrary_write]
  assurance:
    replayability: [none, trace_only, deterministic_replay]
  prohibit_understated_exposure: true
  prohibit_overstated_replayability: true

artifact_consistency:
  authoritative_nodes: [criterion_ledger, final_decision]
  invariants:
    - "certificate.decision == recompute(criterion_ledger, policy).decision"
    - "report.score == certificate.score"
    - "completed_dossiers >= delivery_tier.minimum_completed"
    - "all(package.asset_refs).resolve"
  publication:
    freeze_before_hash: true
    verify_registered_bytes: true
    mode: atomic_directory_rename

lineage_closure:
  evidence_identity:
    duplicate_key: [source_identity, content_hash]
    preserve_independent_corroboration: true
\end{lstlisting}

\subsection{Canonicalization, signatures, and content binding}

A SIP-RC manifest MUST state its canonicalization and digest algorithms. A minimal implementation MAY use JSON Canonicalization Scheme and Ed25519, as in the local prototype. Production profiles SHOULD reuse established content-binding and provenance conventions where applicable, including C2PA collection bindings or equivalent signed asset lists. The signature MUST cover the manifest fields that determine authority, policy, relations, versions, and registered artifacts.

Revocation MUST distinguish a historically valid release from a release that never conformed. A changed registered artifact invalidates the release; it cannot be patched without rebuilding and re-verifying the manifest.

\subsection{Validation and publication algorithm}

\begin{scriptsize}
\begin{verbatim}
handshake and verify profile
  -> validate object schemas
  -> validate semantic and operator bindings
  -> close claim evidence and lineage
  -> recompute decisions on a separate path
  -> validate scope and completeness gates
  -> render candidate views
  -> validate view, asset, and package relations
  -> freeze registered files
  -> hash, sign, and read-only re-verify
  -> publish atomically or reject
\end{verbatim}
\end{scriptsize}

An accepted package MUST be byte-identical to the verified package. A validator MUST emit stable codes, affected nodes/edges, expected and observed values, and permitted remediation. On a blocking violation, it MUST continue evaluating all independent predicates so the release receives one consolidated diagnostic result, except where the failure makes a later predicate semantically undefined or unsafe to execute.

\subsection{Typed human approval}

“Human in the loop” is not one interchangeable mechanism. Formula-fidelity review, outbound-send authorization, visible-UI confirmation, and receiver acceptance approve different risks and artifacts. A required approval MUST bind the decision type, approver role, immutable artifact version, hash, and time. Approval of one artifact or risk MUST NOT be reused as approval of another unless an explicit policy permits that delegation.

\begin{lstlisting}[language=yaml]
human_approval:
  artifact_ref: "report-v3"
  decision_type: "fidelity_review"
  authority_scope:
    semantic: "observation"
    publication: "artifact"
  approver_role: "domain_reviewer"
  artifact_hash: "sha256:..."
  timestamp: "2026-07-13T14:00:00Z"
  decision: "approve"
\end{lstlisting}

This requirement connects directly to Schema Docs, where Send Gate approval, formula review, desktop-visible verification, and receiver acceptance are separate release gates.

\subsection{Partial release}

A release graph MAY declare independently releasable subgraphs. If a node is blocked, every authoritative or rendered descendant MUST also be blocked unless a declared cut removes the dependency and the remaining subgraph is revalidated. Partial release requires a new manifest and hashes. A blocked claim MUST NOT remain visible through a summary, certificate, cached view, or asset caption.

\subsection{Constraint relaxation}

\begin{table}[htbp]
\caption{Classification of constraint relaxation behaviors.}
\label{tab:relaxation}
\centering
\setlength{\tabcolsep}{4pt}
\begin{tabular}{p{1.8cm}p{2.2cm}p{3.2cm}}
\toprule
\textbf{Class} & \textbf{Examples} & \textbf{Allowed response} \\
\midrule
Security-rigid & signature, file/network/tool scope, lease & block or authorized human decision \\
Epistemic-rigid & evidence, operator, mode honesty & collect evidence, mark unknown, or block \\
Procedure-adaptive & tool order, provider route, budget & replan within rigid bounds \\
Expression-soft & layout, serialization, wording & transform or visibly degrade \\
\bottomrule
\end{tabular}
\end{table}

The \texttt{geo\_hello\_sandbox} prototype verifies typed fallback suggestions for soft formatting and rigid file-scope rejection. It does not by itself prove a complete audited retry lifecycle.

For example, a renderer MAY change Markdown layout under an \texttt{Expression-soft} rule, but it may not rewrite a failed certificate as passed. A provider outage MAY trigger \texttt{Procedure-adaptive} rerouting within an approved network scope, but it may not downgrade a \texttt{live-evidence-required} claim to cached evidence without marking the claim unknown or blocked. A human risk exception may authorize a specifically hash-bound release where policy permits; it may not silently relax signature integrity or evidence truthfulness.

\subsection{Representative relational error codes}

A validator MUST use stable error identifiers shared by conformance vectors and incident-corpus labels. The main profile retains representative release-blocking codes; the complete taxonomy is provided in the V15 Supplement.

\begin{table}[htbp]
\caption{Representative relational error codes.}
\label{tab:errors}
\centering
\setlength{\tabcolsep}{4pt}
\begin{tabular}{p{2.4cm}p{2.3cm}p{2.5cm}}
\toprule
\textbf{Code} & \textbf{Meaning} & \textbf{Default action} \\
\midrule
\texttt{\tiny SIP\_RC\_OPERATOR\_MISMATCH} & Operator and status disagree & Block affected claim and descendants \\
\texttt{\tiny SIP\_RC\_CLAIM\_UNSUPPORTED} & Claim lacks evidence or operator & Block claim; revalidate partial graph \\
\texttt{\tiny SIP\_RC\_MODE\_PROMOTION} & Silent strength promotion & Block affected artifact \\
\texttt{\tiny SIP\_RC\_SCOPE\_ESCALATION} & Lower-scope authorizes higher-scope & Block action; preserve bounded implication \\
\texttt{\tiny SIP\_RC\_POST\_FREEZE\_MUTATION} & Registered bytes changed after freeze & Invalidate and rebuild package \\
\texttt{\tiny SIP\_RC\_ASSET\_UNREACHABLE} & Reference does not resolve in package & Block package or affected subgraph \\
\bottomrule
\end{tabular}
\end{table}

\subsection{Threat boundary}

SIP-RC is an information, decision, and artifact-release boundary. It does not contain arbitrary malicious code, prevent kernel exploits, isolate system calls, repair an over-privileged host credential, or recover a compromised operating system. High-risk capabilities still require a subprocess, container, gVisor, microVM, hardware, or equivalent execution boundary. Runtime controls govern what may happen; SIP-RC governs whether the resulting package may be published.

The Supplement carries the detailed threat table and the guidance on evidence minimization, patents, and trade secrets \cite{epo55, uspto, wipo, ecgdpr, mccallister2010pii}.

\section{Implementation Status and Evaluation Plan}

The current evidence stops at implementation feasibility. DRSS supplies observed failures. The seven invariants were derived from those failures and the two other development histories. Schema Docs implements several of them and protects them with regression tests. The profile as a whole has not been compared with the baselines below.

The next study begins by freezing the profile and building a corpus of 40–60 root causes. Internal incidents are split before evaluation: no more than 60\% may shape the vocabulary, up to 20\% may support validator development, and at least 20\% remain blind. One root cause counts once even when it triggers many assertions. A fourth system, maintained outside psi.run and chosen after the freeze, provides the external transfer test.

Two validators will read the same public profile and vectors without sharing decision code. Validator A works over canonical JSON or CBOR. Validator B lifts the graph into SHACL and a rule layer. That separation is architectural; the study will report authorship and organizational control separately.

The principal comparisons are object-schema validation, generic graph validation, provenance/content binding, shared-generator consistency, runtime controls, and the full candidate profile. Ablations remove one path at a time: separate-path recomputation, risk/assurance propagation, rendered-artifact checks, scope enforcement, mandatory gates, or atomic publication verification.

Primary outcomes are relational-fault recall, false blocks, escape rates, validator–human agreement, validation latency, and contract-authoring cost. The analysis will emphasize paired incident-level outcomes, effect sizes, and confidence intervals; null results will not be treated as equivalence. The corpus schema, baseline definitions, ablation matrix, statistical plan, reproducibility inventory, and confidentiality fields are specified in the Supplement.

Until those results exist, performance, false-block rate, and cross-implementation interoperability remain unanswered.

\section{Discussion}

The cases did not call for a larger schema. They changed the object under review. The object is now the release: its derivation, rendered views, approvals, and final bytes.

\subsection{Why Schema Docs is the principal success case}

Schema Docs changed what failure means inside the product. A conversion can lose fidelity, require review, or stop. Those states follow the document through formulas, assets, masking, audience views, packaging, receiver checks, and external editing. The regression suite matters because it preserves failures found in real files, not because its total is large. A skipped environment-dependent test stays skipped, and an unverified desktop fixture stays blocked. Loss is acceptable when the product names it; silent success is not.

\subsection{Decision scope as an authority boundary}

Security models normally govern identities, tools, files, and networks. DRSS shows that authority also exists inside a reasoning product. A criterion-level observation can overreach into a project-level action without invoking an unauthorized tool. Decision-scope enforcement treats implication as a capability: lower-scope claims may influence higher-scope decisions only through declared operators.

\subsection{Provenance is necessary but not a release decision}

PROV-O, in-toto, and C2PA offer mature representations and bindings that SIP-RC should reuse. A perfectly traceable contradiction remains a contradiction. SIP-RC adds the policy that certain provenance-bearing relations are release-blocking and identifies agent-specific predicates for claim, decision, view, tier, and publication agreement.

\subsection{Refutability rather than omniscience}

SIP-RC cannot decide unrestricted factual truth. It can expose the basis on which a package asks to be trusted. An auditor can then point to ineligible evidence, an inverted operator, an overreaching claim, promoted execution conditions, a missing dossier, or bytes changed after review. This makes release relations auditable within the broader governance, technical-control, and deployment context described by NIST AI RMF \cite{nist2023airmf}.

\section{Limitations}

\subsection{Common-source bias}

All three systems originate from one research program. Antigravity provides a separate development account but not an external replication. The proposed vocabulary may reflect local terminology and architecture.

\subsection{Retrospective collection}

The histories were not preregistered. Some incidents are reconstructed from development logs and acceptance tasks. Implementation status is therefore explicit, and pending tasks are not counted as completed repairs.

\subsection{Unequal maturity and test semantics}

Schema Docs and GEO have freshly executed current suites. DRSS supplies a historical artifact counterexample. The suites test different risks, so Table 2 reports coverage categories rather than treating totals as comparable evidence.

\subsection{Manifest and ontology quality}

SIP-RC can only enforce declared semantics. An incomplete or strategically permissive manifest may validate a non-conformant real-world method. ChainCaps similarly identifies manifest quality as a deployment bottleneck \cite{jiang2026chaincaps}. Human review, negative vectors, schema governance, and cross-system reuse tests are necessary to distinguish profile quality from validator correctness.

\subsection{Validator fallibility and monoculture}

A validator may implement the declared invariant incorrectly. A local canonicalization prototype previously accepted an unpaired Unicode surrogate until a focused test exposed the bug. If generator and validator share the same assumptions, apparent conformance may reproduce a common error. Cross-language or differently modeled implementations, adversarial vectors, and human adjudication are required.

\subsection{Evaluation status and generalization}

The current study is classified as a design and experience contribution. The evaluation protocol is intended for preregistration; once run, it will test whether the observed mechanisms generalize to a broader incident class and whether SIP-RC provides detection advantages over the specified baselines at acceptable validation and authoring cost.

\subsection{Formal scope}

The conditional conformance guarantee depends on a declared fault set and correct validator implementation. It is a target, not a theorem, and makes no claim about open-world truth, scientific validity of the domain method, or absence of undeclared failures.

\section{Conclusion}

DRSS passed every local gate and still shipped a package whose ledger, certificate, report, and audience claims disagreed. Schema Docs shows the opposite trajectory: failures found in real documents became contracts over content, views, assets, caches, approvals, and receiver checks. GEO shows that some of the same relations appear when evidence leads to work rather than a report.

SIP-RC gives those relations a form that a validator can test: support, authority, execution risk and assurance, lineage, cross-artifact agreement, hard completeness gates, and freeze–verify–publish integrity. That is the contribution reported here. The comparative question now belongs to the corpus, the two validators, the blind hold-out, and the external system.

\bibliographystyle{IEEEtran}

\end{document}